\documentclass{aa}  

\usepackage{natbib,graphicx,hyperref}
\usepackage{txfonts}
\begin{document}

   \title{Towards a better coordination of Multimessenger observations}

   \subtitle{VO and future developments}

   \author{J.-U. Ness\inst{1}
          \and
          C. S\'anchez Fern\'andez\inst{1}
	  \and
          A. Ibarra\inst{1}
          \and
          R. Saxton\inst{1}
          \and
          J. Salgado\inst{1}
          \and
          E. Kuulkers\inst{2}
          \and
          P. Kretschmar\inst{1}
          \and
          M. Ehle\inst{1}
          \and
          E. Salazar\inst{1}
          \and
           C. Gabriel\inst{1}
	   \and
	  M. Perdikeas\inst{3}
          }

   \institute{European Space Astronomy Centre (ESAC) at European Space Agency (ESA)\\
              \email{juness@sciops.esa.int}
         \and
	 European Space Technology Centre (ESTEC) at European Space Agency (ESA)
	 \and
	 Chandra X-Ray Center Data Systems, Smithsonian Astrophysical Observatory, Cambridge, MA 02138, USA
             }

   \date{Received \today, 2018; accepted }

 
  \abstract
   {Towards the multimessenger era, the scientific demand for
   simultaneous observations with different facilities is continuously increasing. The main
   challenges of coordinating observations
   is the determination of common visibility slots and the respective
   scheduling constraints to find common free slots. While all
   this information is publicly available via the respective
   observatory web pages, it is cumbersome to find this information
   from a large diversity of web interfaces and web tables.}
   {While coordinated observations have been planned already in the
   multiwavelength area, their number will increase,
   and the larger complexity requires much better use of automatic
   routines.}
   {Automatic tools are not able to obtain visibility and planning
   information from web interfaces, and it is thus necessary to
   develop standard interfaces between observatories and automatic
   tools. We have developed two Virtual Observatory (VO) protocols
   ObjVisSAP and ObsLocTAP that work with a URL-based query approach
   with standardized query parameters and standardized output.
   Clients can then pull the required information directly from the
   respective observatories using the visibility and observation
   locator services and compute overlapping, unplanned, visibility
   intervals. Many other use cases are possible.}
   {A prototype service has been implemented by the INTEGRAL mission
   and a very simple client script queries visibility intervals for
   given coordinates within the next 30 days. An observer can then
   quickly see whether a source is observable within the near
   future. Other missions are on their way to implement
   the services.}
   {Once the major observatories have implemented the
   services and powerful tools are available that rely on getting
   visibility and planning observations via these protocols, we
   expect many other, also ground-based, observatories to follow.
   We are developing documentation to support observatories with
   the implementation.
   }

   \maketitle
%

\section{Introduction}

Astronomical observatories individual ranges of scientific investigations but no single observatory can satisfy all needs. Depending on scientific goals, data from different observatories have to be combined while strictly simultaneous observations are not always required. The variability time scale of the source in the considered wavelength bands dictates how much separation between observations can be tolerated.\\
From the perspective of XMM-Newton operations, for example, the demand for coordinated observations has increased during the last decade.
\begin{figure}[!ht]
\resizebox{\hsize}{!}{\includegraphics{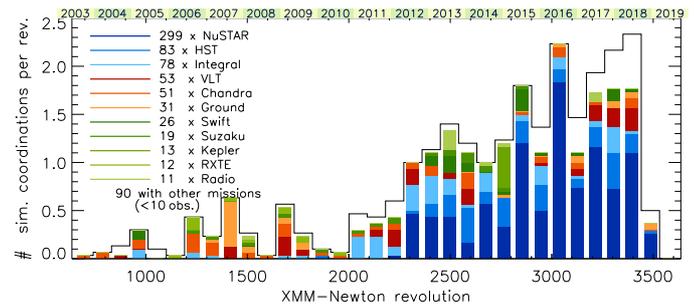}}
	\caption{\label{coordevol}Evolution of number of observations that have been performed simultaneously with XMM-Newton and other missions. The horizontal time axis is XMM-Newton revolution (2 days) with rev0 being 10 December 1999. In the top, the corresponding years are given.}
\end{figure}
In Fig.~\ref{coordevol}, we show the annual evolution of the number of XMM-Newton observations that were taken simultaneously with other space- and ground-based observatories. Until $\sim 2010$, the XMM-Newton schedule contained one coordinated observation about every four revolutions ($\sim 1$ week) while $\sim 2-3$ observations are on average scheduled per revolution. With the launch of NuSTAR, the demand for coordinated observations increased rapidly. A further increase in coordination can be expected with additional joint programmes plus big campaigns related to multi-messenger follow-ups.\\

\begin{figure}[!ht]
	\resizebox{\hsize}{!}{\rotatebox{270}{\includegraphics{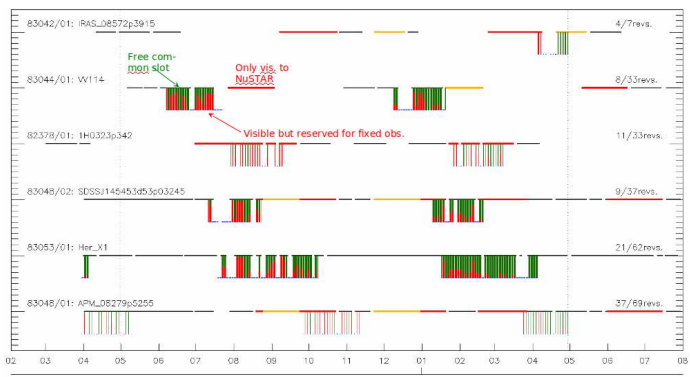}}}
\vspace{.1cm}
	\caption{\label{mcoord}Illustration of semi-automatic coordination
	between XMM-Newton and NuSTAR. The horizontal axis is time in months
	of the years 2018/19 and in vertical direction six example targets
	are given that need simultaneous XMM-Newton/NuSTAR observations.
	The horizontal lines next to each target indicate visibility
	to NuSTAR (without Earth blocks) with red/orange/black colour
	indicating highest to lowest preferences.
	XMM-Newton visibility is indicated by horizontal blue bars.
	Common visibility slots are indicated by vertical bars between the
	respective NuSTAR and XMM-Newton visibility lines, where green
	indicates free slots and red that XMM-Newton had already planned
	observations. The numbers to the right indicate the number of
	unblocked common visibility and total number of common visibility
	slots. The visibility information was provided by the respective
	missions.
	}
\end{figure}

The scheduling of coordinated observations between different observatories
is usually done via e-mail or telephone communications between the respective
scientific planning teams. A controllably low number of coordinated observations allows
the teams to coordinate each target individually. However, for XMM-Newton's
17th observing cycle (AO17), the number of coordinated observations was already close to 30 (not
counting triggered observations). The increasing number and complexity of
coordination requirements
demands the use of more advanced approaches to the problem, such as the use
of optimization techniques that account for the various coordination
requirements, explore the possible solutions and provide an optimized
observation plan. Automatic planning optimization techniques are already
used by, e.g., INTEGRAL or Swift \citep{aaron} but they only work for a single
facility with direct access to mission-internal information. Multi-facility tools
require automatic access to visibility and planning information from all
involved observatories.\\

In order to handle the larger number of AO17 XMM-Newton observations,
an internal tool was developed to aid the manual process. Out of the pool
of targets, some are easy, others are more complicated to coordinate,
depending on the amount of overlapping visibility intervals that are
still available, thus not yet blocked by high-priority time-constrained
observations. The tool automatically determines the common visibility slots
plus the respective subsets of slots that are already blocked by
time-constrained observations. It then produces a graphical illustration
(see Fig.~\ref{mcoord} for six targets) and determines the best order in which
to arrange the coordinations with NuSTAR, thus starting with the target
with the smallest number of possibilities (top line). Generally,
the first free common slot is selected in order to leave margins for alternatives,
for example if a coordinated observation has to be replanned to accommodate a
triggered observation.\\

At this stage, the tool depends on the respective visibility and planning
information to be given, and the idea of this project is that the tool can
extract this information independently. This not only reduces the number
of human activities, it also allows the tool the use of the most recent
information.\\

More professional and sophisticated approaches are certainly desirable and are
under development by several groups, e.g., the ASTERICS \citep{asterics}
and Smartnet \citep{smartnet} projects. We focus here on the fundamental
building block: How to obtain the visibility information and the observing
plans from each involved facility.\\

While visibility information is available via web tools, published on the
web sites of each observatory, input and output formats are not standardized
and are subject to change. Planning information is only provided
by some observatories in the form of HTML tables, again with no static nor
standard formats. Under these circumstances it is in theory possible to develop
an automatic way to extract visibility and planning information for each
facility, but it will be extremely unstable and very observatory-dependent.\\

Our group has therefore embarked on a project to define international
standard protocols in which observatories provide information of
object visibility and observation status including performed (past)
and planned (future) observations \citep{kuulkers19}. We have iterated
an initial definition
document with around 60 partners (especially observatories, developers of tools,
scientists) and presented the consolidated version to the Virtual Observatory
with the ultimate goal to obtain VO certification; for VO protocol descriptions, see Reference list.

\section{The proposed standard protocols}
We propose two different protocols based on already existing International
Virtual Observatory Alliance (IVOA) standards. The protocol to obtain the
observation information will be based on the Tabular Access Protocol (TAP)
and is called "Observation Locator Table Access Protocol (ObsLocTAP)". For the
visibility information, a Simple Access Protocol (SAP) will be used because
on the observatory side, tabulated visibility information may not be available
within a database (this being a requirement for the TAP). The visibility service
is called "Object Visibility Simple Access Protocol" (ObjVisSAP).
Both services use a URL-based (REST-like) interface admitting queries with a
\texttt{keyword=value} format returning a table in the VO standard XML format. The
TAP also allows a more powerful query language like ADQL \citep{ADQL}.

\begin{figure}[!ht]
\resizebox{\hsize}{!}{\includegraphics{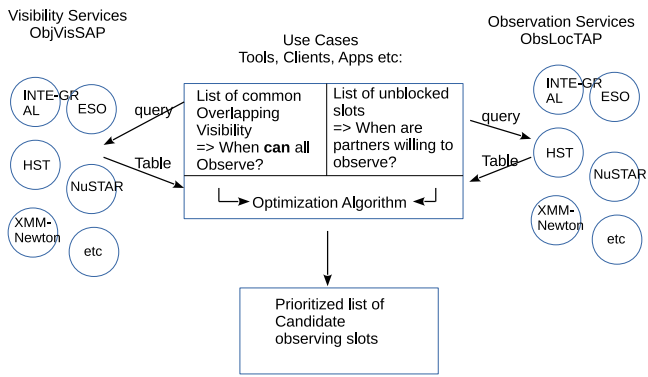}}
\vspace{-1.4cm}
\caption{\label{sketch}Illustration of information flow in the proposed two protocols querying visibility/observing information via identical query calls from each observatory service, processing the information to planning possibilities accounting for hard constraints (left) and soft constraints (right), leading to a consolidated list of prioritized candidate observing slots.}
\end{figure}
\subsection{Target Visibility}
\label{vis}

Query parameters for target visibility need to specify a sky area (e.g.
coordinates, field of view) and an observing time window. It was discussed
whether coordinates and time intervals could be optional input parameters.
We decided that for a compliant service, both time intervals and coordinates
must be implemented by observatories while time intervals may be implemented
as optional parameters with sensible defaults (e.g., start time as current
time).\\
Currently, the respective observatory web sites provide dedicated
visibility tools with web entry fields where users can indicate their
preferences. The output is then provided either as a table
(e.g. HTML or ASCII) with visibility intervals or a
diagnostics plot from which the various constraints can be extracted by visual
inspection. Even when the output is available in some parsable, in principle,
format (such as HTML or ASCII) neither format is suitable to attaching
semantic meaning to it in a standardized and tool-supported way and
thus neither method is amenable to further automated processing. At
best, ad hoc and brittle web scraping is the only automation option.\\
In the proposed standard protocol, the target visibility service's
response is served in XML or JSON formats. Both of these formats are
more suited to structured representation of information without any
presentation-oriented markup (such as the one employed by HTML tables
for example). Moreover, the form of the requests (HTTP request
messages) used to elicit those responses has semantic transparency
(being REST-like) and allows such queries to be scripted and
fine-tuned to extract the exact information needed. Both these
features facilitate automated discovery and further processing of
target visibility information in some pipeline such as the one
suggested by Figure~\ref{sketch}. The client is in the middle, making
standardized queries to visibility (left) and observation (right)
services to many observatories which return tabulated results which
are then processed to yield the desired result, e.g., a prioritized
list of candidate observing slots.\\

The following example script {\tt visibility.sh} in {\tt bash} takes a pair
of sky coordinates as input and returns visibility intervals for INTEGRAL
for the next 30 days as output:\\

\noindent
{\tt \small
\#!/bin/bash\\
ra=\$1\\
if test x\${2} = x; then\\
\indent   echo Need to provide coordinates in decimal units\\
\indent   exit 1\\
else\\
\indent   dec=\$2\\
fi\\
root="http://integral.esa.int/isocweb/tvp.html"\\
dstart=\$(date +\%d-\%m-\%Y)\\
dend=\$(date -d "+30 days" +\%d-\%m-\%Y)\\

\noindent
curl -s "\${root}?startDate=\${dstart}\&duration=12.600\&\\
\indent action=predict\&endDate=\${dend}\&coordinates=equatorial\&\\
\indent ra=\$ra\&dec=\${dec}\&format=json"\\
\indent | jq '.INTERVALS[].revolution'\\
\indent | cut -d'"' -f2 > temp.rev\\
\noindent
curl -s "\${root}?startDate=\${dstart}\&duration=12.600\&\\
\indent action=predict\&endDate=\${dend}\&coordinates=equatorial\\
\indent \&ra=\$ra\&dec=\${dec}\&format=json"\\
\indent | jq '.INTERVALS[].start'\\
\indent | cut -d'"' -f2 > temp.start\\
\noindent
curl -s "\${root}?startDate=\${dstart}\&duration=12.600\&\\
\indent action=predict\&endDate=\${dend}\&coordinates=equatorial\&\\
\indent ra=\$ra\&dec=\${dec}\&format=json"\\
\indent | jq '.INTERVALS[].end'\\
\indent | cut -d'"' -f2 > temp.end\\

\noindent
n=\$(wc -l temp.rev | cut -d' ' -f1)\\
d=\$(date +\%y-\%m-\%d)\\
echo "Today: \$d"\\
for ((i=1; i<n; i++)); do\\
\indent    a=\$(sed -n "\${i},\${i} p" temp.rev)\\
\indent    b=\$(sed -n "\${i},\${i} p" temp.start)\\
\indent    c=\$(sed -n "\${i},\${i} p" temp.end)\\
\indent    echo " \$a   \$b   \$c"\\
done\\
rm temp.rev temp.start temp.end\\
}

Executing this script on February 8th, 2019 with coordinates of Sgr A*:\\

{\tt \small
\noindent
./visibility.sh 266.4 -29\\
}

yields as output a table with columns INTEGRAL revolution,
UT time of visibility start and visibility end, respectively:\\

{\tt \small
\noindent
Today: 08-02-2019\\
 2058   2019-02-17 02:43:20 GMT   2019-02-19 06:51:53 GMT\\
 2059   2019-02-19 18:32:09 GMT   2019-02-21 22:39:52 GMT\\
 2060   2019-02-22 10:19:50 GMT   2019-02-24 14:28:47 GMT
 \vspace{-.2cm}
 \[...\]
}
 \vspace{-.2cm}

In this example, a scientist will see that INTEGRAL can
only observe Sgr A* after February 17.
This script can easily be expanded to probe visibility to
other observatories, only by replacing the root URL.\\

In this test version, we have used the more convenient
json format, but all implementations will be required
to return output in the VO-compliant XML format (VOTABLE).\\

Clients can build more complex applications based on such automatic queries, for example to create a plot such as the one in Fig~\ref{mcoord} without having to consult other planning teams.

\subsection{Observations}

Minimum observing information consists of target coordinates and observation start and end times. Specific additional information such as instrumental setup may be of interest as well. Past and future observing information is not commonly found in a single place. Information of past observations are usually accessible via the respective observatory science archives where also data can be retrieved. Meanwhile, not all observatories publish their observing plans, some only publish a short-term plan while others publish both, short- and long-term plans. These are currently provided in the form of HTML tables in the web that can in principle be read by robots but any changes in format requires re-design of any multi-mission planning tools, making them unstable in the long run.\\
While the target visibility query allows users to determine when any given observatory {\em can} observe a given target, what is really of interest for coordination activities is when observatories of interest {\em are willing} to actually observe it. In addition to the (hard) visibility constraints, certain times may be blocked by high-priority observations for one or more of the involved observatories that could only be cancelled at high scientific cost. Times can also be blocked by other constraints that are not delivered by the visibility server, e.g. special operations, ground station gaps, maintenance. As shown in \S\ref{vis}, it is relatively easy to determine a list of overlapping times of target visibility, but there will be no preferences in such a list. With additional planning information, an optimization algorithm could also rank the visibility time intervals by overall impact on existing observing plans that can then be minimized.\\
While the prioritization is not an analytic function, the observing plans give some elements of an initial assessment. Direct interactions may still be needed, but if the initial concepts are already close to reality, the process of coordinating multiple targets with multiple observatories could be made much more efficient.

\section{Use Cases}

An important element to obtain approval from the IVOA, are use cases that demonstrate that the proposed standards are of common use. Some initial examples are given here.

\subsection{Science Planning}
 \begin{enumerate}

\item Long-term planning of a large number of observations for
simultaneous execution with various observatories. Communication with mission planners is still needed, but conversion is much faster if:
    \begin{itemize}
       \item Visibility services are used for 
       the observatories to find a list of common visibility slots
       \item The long-term plan information
       for the various observatories is taken into account to identify time intervals that are not planned yet.
       \end{itemize}
  For each target, a list of time intervals when all observatories can observe and when they are also free can be generated.
\item Coordinate a fast-response Target of Opportunity campaign
    \begin{itemize}
    \item Use visibility services to get list of common visibility slots
       \item Priorities of planned observations in each involved observatory can be used to rank the common visibility slots to minimize scientific impact at least to those observatories providing priority information on planned observations.
    \end{itemize}
 On short notice, no free slots can be expected, but if priority flags are provided, observations that are flexible (and can thus be postponed) can be identified to find potentially feasible slots.

\item Schedule multi-band observations with diverse constraints (e.g. ground-based only during night, space-based only during certain positions in the orbit) at a fixed orbital period of a system: Visibility information from above can be folded with the ephemeris of the target to find out whether there is any coinciding times the desired phase can be covered by all facilities. If more than one possibility is found, planning information can be used to rank them.

\item A small observatory (or even private garden observatory) wants to follow observations of a large observatory without interaction: The ObsLocTAP service gives the information of current and scheduled observations.

\end{enumerate}
\subsection{Science Exploration}

\begin{enumerate}
\item A scientist is interested in one particular target and wants to plan new observations without duplicating existing efforts: A query on target coordinates without any time limits to all participating observatories gives a list of performed and planned future observations giving an idea of the gaps that need to be filled.

\item A scientist is analyzing data of one observatory and wants to know whether other observatories observed the same target, possibly even during the same time, e.g., silently followed: A multiple query on past observations of all participating observatories, filtered on target coordinates (and time interval if contemporaneous data are desired), gives this information.

\item Is there any gamma ray emission from Saturn? Based on the ephemeris information of Saturn over the mission life time of a given mission, e.g., INTEGRAL, perform multiple ObsLocTAP queries for the respective coordinates and times to find out which INTEGRAL observations had Saturn in the field and return the respective observation identifiers.

\item A scientist wants to find all observations of Jupiter that were taken simultaneously with a radio and an X-ray telescope: A tool that takes the planet ephemeris as input makes queries for short consecutive time intervals and the respective sky coordinates to the radio and X-ray telescopes of interest. Positive output is recorded and simultaneous coverage can then easily be identified.

\end{enumerate}

\section{State of the art and next steps}

During a workshop on Visibility and Observation Locator Protocols held
2018 September 21, more than 60 participants expressed great interest in
the initiative. To convert this interest into concrete action, the following
has been done:
\begin{itemize}
 \item Protocol descriptions have been written and published under IOVA
	 for ObjVisSAP:\\
http://www.ivoa.net/documents/ObjVisSAP/
and ObsLocTAP:\\
http://www.ivoa.net/documents/ObsLocTAP/
 \item Presented to VO
 \item Prototypes for INTEGRAL were implemented
 \item Collaboration with the Chandra X-ray Centre (CXC) who are pioneering the implementation
 \item Developed simple demonstrator scripts to show how to use services
\end{itemize}

\noindent Next steps will be to
\begin{itemize}
 \item Write documentation how observatories can implement the services
 \item Liaise with more observatories to implement the services
 \item We plan to team up with many observatories to publish a refereed article describing the services in a less technical manner than VO protocols and to describe the individual challenges of some observatories.
\end{itemize}
An implementation manual can only cover general aspects to serve as a startup guide while the large diversity of demands from facilities need to be addressed individually. For example, low-Earth orbit space missions cannot provide accurate visibility information for more than 10 days in advance. Since Sun/Moon constraints alone are already of interest, separate short-term and long-term services may be implemented, including and excluding Earth constraints, respectively. The start up guide will thus only describe the implementation of common features based on the experience from INTEGRAL and Chandra.\\
For the refereed publication we envisage to give descriptions of individual characteristics of various observatories. Future observatories may then identify the most similar case to their situation from which they can learn.


\bibliographystyle{aa}
\bibliography{jness}
\noindent
Observation Locator Access Protocol:\\
\href{http://www.ivoa.net/documents/ObsLocTAP/}{http://www.ivoa.net/documents/ObsLocTAP/}\\
\noindent
Object Visibility Access Protocol:\\
\href{http://www.ivoa.net/documents/ObjVisSAP}{http://www.ivoa.net/documents/ObjVisSAP}\\
%

\end{document}